\documentclass[linenumbers]{aastex631}

\usepackage{pythonhighlight}
\usepackage{hyperref}

\nolinenumbers
\begin{document}
\nolinenumbers
\title{Pipeline Provenance for Analysis, Evaluation, Trust or Reproducibility}

\author[0000-0002-5566-6147]{Michael A. C. Johnson}
\affiliation{Max Planck Institute for Radio Astronomy}
\affiliation{German Aerospace Center (DLR), Institute of Data Science}
\affiliation{University of Manchester}

\author[0000-0002-0648-2704]{Hans-Rainer Kl\"ockner}
\affiliation{Max Planck Institute for Radio Astronomy}

\author[0000-0002-2282-5105]{Albina Muzafarova}
\affiliation{BETTA Security GmbH}

\author[0000-0002-6554-3722]{Kristen Lackeos}
\affiliation{Max Planck Institute for Radio Astronomy}

\author[0000-0003-1361-7723]{David J. Champion}
\affiliation{Max Planck Institute for Radio Astronomy}

\author[0000-0002-8180-1525]{Marta Dembska}
\affiliation{German Aerospace Center (DLR), Institute of Data Science}

\author[0000-0002-0964-4457]{Sirko Schindler}
\affiliation{German Aerospace Center (DLR), Institute of Data Science}

\author[0000-0002-5743-6580]{Marcus Paradies}
\affiliation{Technical University of Ilmenau}

\begin{abstract}

\nolinenumbers

\noindent Data volumes and rates of research infrastructures will continue to increase in the upcoming years and impact how we interact with their final data products. Little of the processed data can be directly investigated and most of it will be automatically processed
with as little user interaction as possible. Capturing all necessary information of such processing ensures reproducibility of the final results and generates trust in the entire process. 

\noindent We present  {\tt PRAETOR}\footnote{{\tt \bf P}ipeline p{\tt \bf R}ovenance for {\tt \bf A}nalysis, {\tt \bf E}valuation, {\tt \bf T}rust {\tt \bf O}r {\tt \bf R}eproducibility}, a software suite that enables automated generation, modelling, and analysis of provenance information of Python pipelines. Furthermore, the evaluation of the pipeline performance, based upon a user defined quality matrix in the provenance, enables the first step of machine learning processes, where such information can be fed into dedicated optimisation procedures.  

\end{abstract}


\section{Introduction} \label{sec:intro}

%
%

%
%

%
%
%
\noindent In a general sense, provenance information documents the history of a thing whereas in case of data products, this concept has been adopted to document the entire production history of the item itself.
However, collecting data, generating science-ready datasets and scientific results is a complex procedure, including a broad range of performance measurements of the infrastructure, expert knowledge, and domain-specific aspects in the data processing. Therefore, obtaining provenance that describes which additional information went in and is available on data products is a key factor to generate confidence and trust of their scientific usage.
Here we present \href{https://gitlab.mpcdf.mpg.de/PRAETOR/prov-PRAETOR_public}{\tt PRAETOR}, a software suite, that records provenance and provides a framework to explore such information.\\

\noindent {\tt PRAETOR} has been developed having astronomical use cases in mind \citep{johnson2021astronomical}, but is applicable to a broad range of applications. Astronomical observations are comprehensive examples of taking data: raw and uncalibrated data are transferred into physical meaningful units based on metadata of technical equipment and observatory conditions; cleaning up the (meta)data from spurious values; validating data properties and quality assessment of data products; transformations into science-ready data products; and finally, analysing and validating the scientific results.
Such workflows are mostly implemented in individual steps with semi-automatic processing, but with upcoming data rates there is a need for fully automated systems with self-regulated processing and optimisation of production lines for science-ready data products.

\noindent {\tt PRAETOR} has been developed to collect provenance of workflows (in the following we refer to pipelines as automated workflows), a user interface ({\tt UI}) that allows browsing through the provenance data, a database solution to query and analyse the captured provenance, and the ability to concatenate provenance information of individual pipeline executions.

\section{Software}\label{sec:soft}

%
%
%
%
%
%
%
%
%

\noindent In the development of {\tt PRAETOR}, we have investigated a way to capture provenance from Python based pipelines and developed diagnostic tools to evaluate the pipeline by utilising its provenance information.
To fully understand the inner workings of the astronomical pipeline described above, one needs information on its input parameters, the original dataset, the observatory metadata, the individual function calls and their parameters, and the relation between all of these pieces of information. Such a set of information is a complex structure and extracting these details is a challenging task.\\

\noindent In order to document this information, we extended the standard language for recording provenance (\href{https://www.w3.org/TR/prov-overview/}{PROV}) and its data model for things, processes, and those responsible for processes as entities, activities, and agents, respectively \citep{belhajjame2013prov}.
However, the main extensions to {\tt PROV} were in the form of attributes and were defined to represent specific components that are common within Python pipelines such as: function names, Python modules and versions thereof, and memory consumption of individual processes. 
For analysis purposes, we implemented a quality metric attribute which can be attached to any component within the pipeline as well as to the pipeline itself.
Based on the astronomical use case analysis \citep{johnson2021astronomical}, we have identified a number of items to extend the \href{https://praetor.pages.mpcdf.de/prov-PRAETOR\_public/}{\tt PROV model for PRAETOR}, but their implementation will be left to future work.
Many of these extensions are relations between different objects within the provenance, e.g. whether an object was used as data or as parameter, or whether a process was responsible for loading a specific object.
However, the motivation for using a minimally adapted {\tt PROV} in the current release of {\tt PRAETOR} was compatibility with existing tools designed for {\tt PROV}, such as the \href{https://github.com/lucmoreau/ProvToolbox}{ProvToolBox} and \href{https://github.com/DLR-SC/prov2neo}{prov2neo}.
In addition, the generated {\tt PRAETOR}-based provenance is designed to be interoperable with tools developed for other {\tt PROV}-based provenance models, such as those from IVOA \citep{servillat2020ivoa}.\\


%
%

\noindent Extracting provenance information and the relation of the individual operations can be an overwhelming task, in particular, if the pipelines are unknown and treated as "black boxes".
In order to obtain a first overview of the available information a {\tt UI} has been developed to browse through the provenance data. 
Once a deeper understanding of the available information has been acquired, queries and analysis on the captured provenance can be done via database operations. 

The {\tt UI} provides a general overview of the used software packages, tools, files, the processing time, basic memory consumption, and individual functions. 
Information of the individual functions can be accessed on a dedicated page that provides details on invocations of functions in the pipeline and their in- and out-put parameters. To investigate the sequence of function calls within the pipeline itself adjacent activities can be followed.

Apart from being a complex structure, provenance data can also grow to substantial sizes and efficient processing can be a limiting issue.
Therefore the {\tt PRAETOR} package provides a framework for uploading provenance to two different kinds of database structures -  graph databases \href{https://neo4j.com}{Neo4j} and triple stores \href{https://jena.apache.org/documentation/fuseki2/}{RDF/fuseki}.
A set of queries are available which extract key information into {\tt pandas} dataframes, such as: function invocations, inputs, and outputs.\\


\section{A first glimpse}\label{sec:n2nexample}

\noindent The software suite has three main components: provenance generation/capturing, provenance analysis and the {\tt UI}. 
The generation and analysis components are packaged within the same Python package, whereas the {\tt UI} is deployed as a series of Docker containers.
The Python package is available on the \href{https://libraries.io/pypi/praetor}{\tt pypi-hub} and can be installed using e.g. {\tt pip install praetor} in either a virtual environment or container alongside an existing pipeline.
The information captured by the package includes any imported modules, functions, file access, and other core functionality to Python.
Which of these pieces of information is included can be defined in the {\tt praetor\_settings\_user.py} settings file.
The {\tt UI} can be installed by cloning the \href{https://gitlab.mpcdf.mpg.de/PRAETOR/prov-PRAETOR\_public/-/tree/main/prov-PRAETOR/user\_interface}{gitlab repository} and following the relevant installation instructions to start up its Docker environment. Once running, the {\tt UI} will be available via the local host in an web browser and the queries are based on to the triple store queries within the {\tt PRAETOR} analysis package.\\

\noindent Various tutorials are available including a full \href{https://gitlab.mpcdf.mpg.de/PRAETOR/prov-PRAETOR\_public/-/tree/main/prov-PRAETOR/provenance\_generation/singularity/prov-PRAETOR\_nutshell\_example}{n2n-example} that explains how to build a Singularity container and generate provenance of an example pipeline.
The example pipeline calls various functions, each having different input and output parameters and are called either within the pipeline itself or via a module.
A subset of the provenance of the example pipeline is shown in Figure~\ref{fig:ui}. This information can also be obtained using the database framework of {\tt PRAETOR}.
For this, we suggest to do the installation within a virtual environment like Conda as explained in the n2n-example.

\section{Conclusion}\label{sec:conc}

\noindent Provenance is of timely importance, by documenting the pathway through scientific processing and by publishing FAIR (Findable, Accessible, Interoperable, Reusable) data products \citep{Wilkinson2016}.
We have presented {\tt PRAETOR} - a software suite for automated generation and analysis of provenance information from Python based pipelines.
We explained the rationale for capturing provenance in a {\tt PROV} extension and the decision to divide the software into three stand-alone packages to operate on any Python pipeline.
A showcase of the analysis of generated provenance via the user interface and the database framework has been presented. 

\noindent The provenance information of multiple executions of a pipeline can be used to characterise the input and output parameters. Pipelines that automatically produce science-ready data sets, often have the problem of data irreversibility, where the raw input data and intermediate results are not stored for repeat analysis. {\tt PRAETOR} provides the tools to evaluate such pipelines and is the basis to any AI and neural network optimisation.

\begin{acknowledgments}
We acknowledge the support of the German Federal Ministry for Economic Affairs and Energy on the basis of a decision of the German Bundestag under the project number 50OO1905.
\end{acknowledgments}

\vspace{5mm}



\bibliography{sample631}{}
\bibliographystyle{aasjournal}

\begin{figure}[b!]
    \centering
    \includegraphics[width=0.15\textwidth]{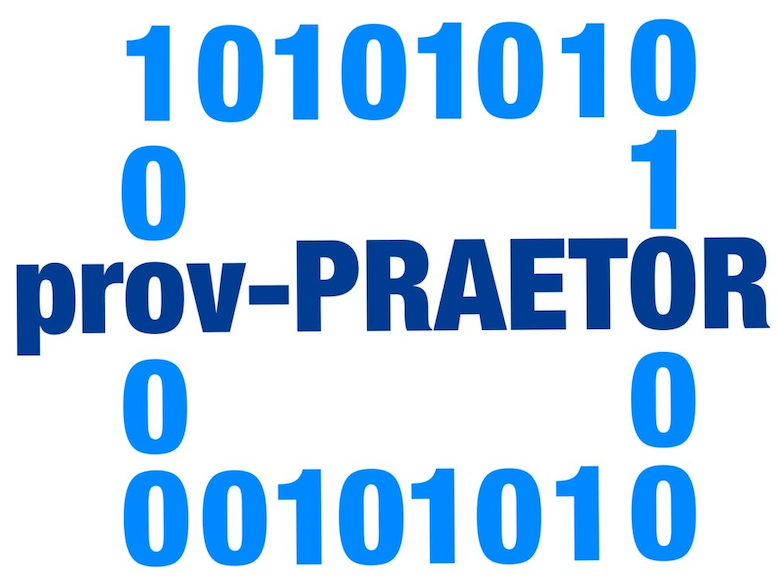}
\end{figure}

\begin{figure}[b!]
    \centering
    \includegraphics[width=0.72\textwidth]{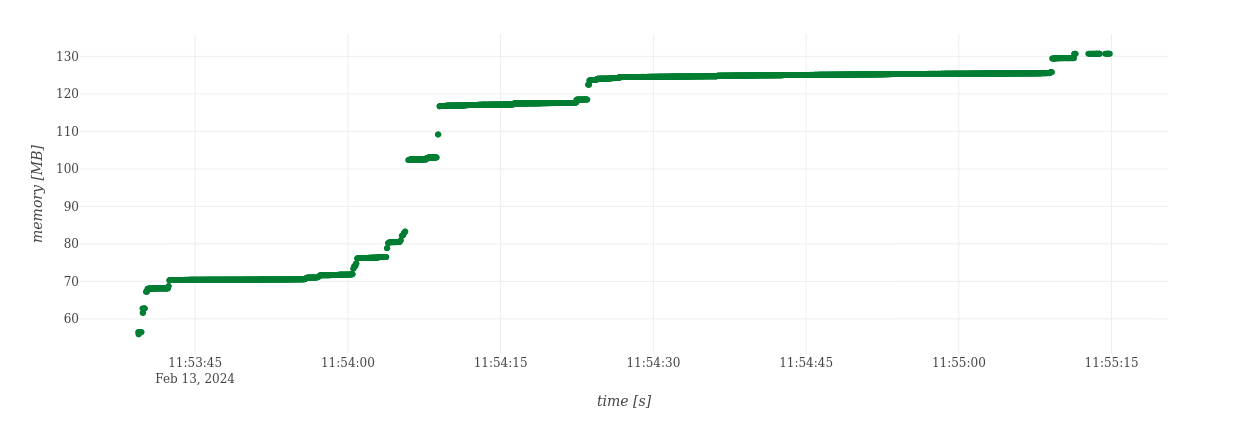}
    \caption{Memory consumption of the processing of the pipeline over time.}
    \label{fig:uioverview}
\end{figure}

\begin{figure}[b!]
    \centering
    \includegraphics[width=0.72\textwidth]{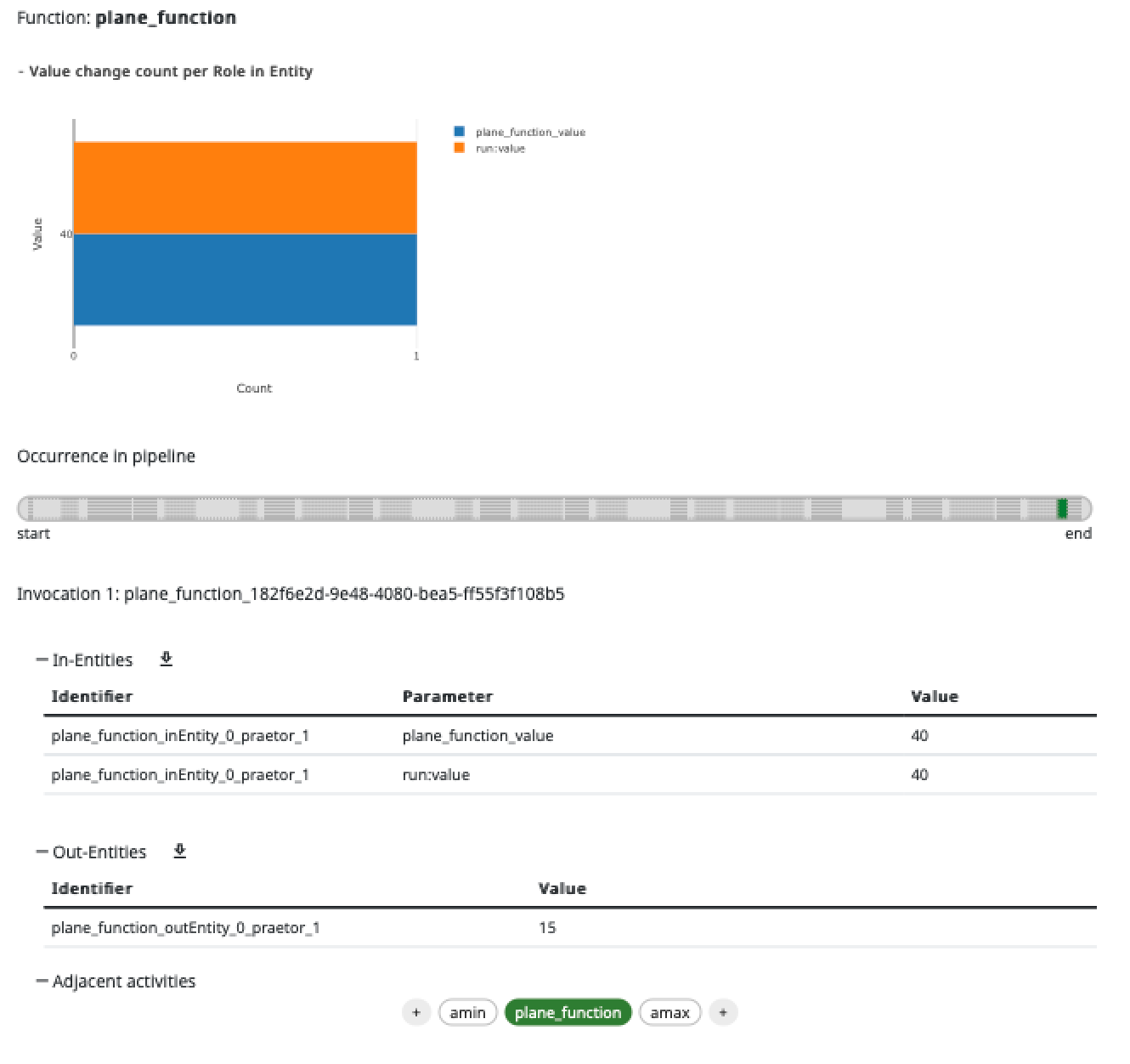}
    \caption{{\tt plane\_function} call, showing the in- and output information and the occurrence of that function in the pipeline.}
    \label{fig:uiinfoplanefunc}
\end{figure}

\end{document}